\documentclass[12pt]{article}

\newcommand{\url}[1]{}

\newcommand{\inverse}[1]{{\textstyle\frac{1}{#1}}}
\newcommand{\half}{\inverse{2}}
\newcommand{\sbar}{\nu}
\newcommand{\sigb}{\bar{\sigma}}
\newcommand{\nub}{\bar{\sbar}}

\title{{\large Banca IMI, PDG internal report}\\ 
{\small (Reduced version in Finance \& Stochastics 4, pp. 147-159, February 2000)}\\
{\bf Discrete Time vs Continuous Time \\ Stock-price Dynamics \\
and implications for Option Pricing}
\thanks{
We are grateful to Aleardo Adotti,
our current head at the Product Development
Group of IMI/San Paolo,
and to Renzo Avesani,
our former head at Risk Management of Cariplo,
for encouraging us in the prosecution of the most speculative
side of research in mathematical finance.
We are grateful also to Wolfgang Runggaldier
and to two  anonymous referees for their remarks and suggestions.
Finally, the paper has been improved both in exposition and
contents thanks to private communications with Prof. Hans
Nieuwenhuis from the University of Groningen} }
\author{Damiano Brigo \hspace{3cm} Fabio Mercurio \\
Product and Business Development Group \\
Banca IMI, San Paolo IMI Group \\
Corso Matteotti 6\\
  20121 Milano, Italy \\
  Fax: 39 02 7601 9324  \\
E-mail: brigo@bancaimi.it
\hspace{1.5cm} fmercurio@bancaimi.it
}
\date{}
\newtheorem{theorem}{Theorem}[section]

\newtheorem{lemma}[theorem]{Lemma}
\newtheorem{corollary}[theorem]{Corollary}
\newtheorem{definition}[theorem]{Definition}

\newtheorem{remark}[theorem]{Remark}
\newtheorem{problem}[theorem]{Problem}
\newtheorem{problems}[theorem]{Problems}
\begin{document}
\maketitle
\thispagestyle{empty}
\begin{abstract}
In the present paper
we construct stock price processes with the same marginal log-normal
law as that of a geometric Brownian motion
 and also with the same transition density
(and returns' distributions) between any two instants in a given
discrete-time grid.

We then illustrate how option prices based on such processes differ
from Black and Scholes', in that option prices can be either arbitrarily close to
the option intrinsic value or arbitrarily close to the underlying stock price.

We also explain that this is due
to the particular way one models the stock-price
process in between the grid time instants which are relevant
for trading.

The theoretical result concerning scalar stochastic differential
equations with prescribed diffusion coefficient whose
densities evolve in a prescribed exponential family, on which part of
the paper is based, is presented in detail.
\end{abstract}

\subsection*{Keywords}
      Stochastic Differential Equations,
      Fokker--Planck Equation,
      Exponential Families,
      Stock Price Models,
      Black and Scholes model, Option Pricing,
      Trading Time Grid, $\Delta$-Markovianity,
      Market Incompleteness, Option replication error.
\section{Introduction}
Since the seminal papers by Black and Scholes (1973) and Merton (1973),
many researchers have tackled the problem of option pricing and hedging
by resorting to continuous-time mathematics to fully exploit the richness
of its theoretical results. In particular, the underlying stock price dynamics
has often been modelled through a geometric Brownian motion. This assumption is still
quite popular, especially among practitioners, even though the empirical distribution of
stock returns is often found more leptokurtic than the normal one.
Indeed, the tradeoff between analytical tractability and empirical
findings still makes the geometric Brownian motion a quite acceptable choice.

However, as we shall prove in this paper, the geometric Brownian motion
is not the only continuous-time process that possesses a lognormal marginal distribution and
normal log-returns. In fact, we shall construct a family of processes with these
characteristics and then we shall also analyse the main implications of the existence of
such a family as far as option pricing is concerned.

In order to achieve our goal, we consider scalar
nonlinear diffusion processes whose densities evolve in
given finite--dimensional exponential families, as from Chapter 7 of Brigo (1996).
We begin by a stochastic differential equation (SDE) a la Black and
Scholes for the stock price, whose solution is the traditional
geometric Brownian motion. The probability density of such a stock
price model evolves in the exponential family of log-normal densities.
It is then possible to define a second SDE with a different (and arbitrarily given)
diffusion coefficient such that its solution has the same marginal
lognormal density as that of the Black and Scholes (1973) geometric Brownian motion.

Furthermore, one can choose the drift of this second SDE in such a
way that its solution has transition densities, between two any instants
of a prescribed finite subset of the time interval, that match the
transition densities of the Black and Scholes process.
We thus derive a family of stock price processes that behave almost
equivalently to a geometric Brownian motion in that they
are probabilistically identical along the finite subset of the time interval.

We then consider the problem of option pricing in a
continuous-time framework. The interesting result we obtain is
that notwithstanding the above mentioned affinity with the Black and Scholes model,
the stock price processes of our family
lead to completely arbitrary option prices, ranging the whole interval of prices
inbetween the no--arbitrage lower and upper bounds.
%
Such a seeming paradox can be explained in terms of the differences between the dynamics of
these stock price processes at an infinitesimal level.

Our result is similar in spirit to that of Rogers and Satchell (1996). However,
our approach is much more constructive in that we are able to provide, under the
original measure, easy and explicit dynamics for the stock price that are
theoretically consistent with whatever option price is observed in reality.

The practical implications of our result can be analysed by considering the case of
a practitioner who has to price an option. Although the physical time he works in is
discrete, he usually resorts to continuous-time mathematics modelling the stock price
underlying the option through a Black and Scholes (1973) process. However,
as the previous theoretical result implies, the geometric Brownian motion
is only one of infinitely many processes that possess the discrete-time properties
required by the practitioner. Therefore, as far as the stock price dynamics is concerned,
the practitioner must view all these processes as equivalent to each other. Yet, when
option pricing is taken into account, he must also be aware that they all imply different,
and arbitrary, option prices. Since this paradoxical situation reveals a clear limit of
the approximation of discrete time with continuous time, our practitioner can react as
follows. He simply avoids the theory of market completeness and just chooses a process
which is consistent with the option price he believes in.
Such an option price can be either exogenously given by the market or
endogenously produced by a model for pricing and hedging in incomplete
markets. This is a further contribution of the paper. We do not derive
option prices and hedging strategies assuming a specific evolution of the
underlying stock price. Instead, we define a family of processes with equivalent
characteristics among which one can select a single process according to his pricing
and hedging purposes.

The paper is structured as follows. In Section 2, the theoretical
result on the construction of some suitable diffusion processes is briefly
reviewed.
Section 3 deals with the application of this result to the case of the
Black and Scholes (1973) model. Section 4 analyses the problem of option
pricing in continuous time and considers a natural particular case.
Section 5 analyses some practical implications of using any of our alternative
processes in describing the asset price dynamics.
Section 6 concludes the paper.

A reduced version of this paper can be found in Brigo and Mercurio (2000).

\section{Stochastic differential equations and exponential families}
In this section we consider the technical result the paper is based on.
This result solves the following problem
for scalar diffusion processes: Given a diffusion coefficient
and a curve in an exponential family of densities, find a drift such that
the solution of the resulting stochastic differential equation (SDE)
has a density evolving in the prescribed exponential family according to the
given curve.
This problem has a straightforward solution.
In this section, we shall present a short summary of the steps leading to the problem
formulation and to its solution.
This summary is based on Chapter 7 of Brigo (1996). A related result appeared in Brigo (1997),
and the general result with its applications both to mathematical finance and
to stochastic nonlinear filtering is also reported in Brigo (2000).

Let us consider the SDE
\begin{eqnarray} \label{sde1}
   dX_t =  f_t(X_t) dt + \sigma_t(X_t) d W_t,
\end{eqnarray}
where $\{W_t, t\ge 0\}$ is a standard Brownian motion independent
of the initial condition $X_0$.
In order to make sure that we are dealing with an equation whose
solution exists unique, we formulate the following assumption.

\begin{itemize}
   \item[(A1)] The initial condition $X_0$ is
      a continuous random variable with density $p_0(x) >0$
      for all $x \in {\bf R}$
      w.r.t.\ the Lebesgue measure on ${\bf R}$ whose moments
      of any order are finite.
      Moreover, the stochastic differential equation~(\ref{sde1})
      characterized by the coefficients $f$, $\sigma$,
      and by the initial condition $X_0$ admits a unique strong solution.
\end{itemize}

Explicit conditions ensuring (A1) are, for example, local Lipschitz continuity
and linear growth, or the Yamada-Watanabe
condition (see e.g. Rogers and Williams (1987), Section V-40).

Once existence and uniqueness of the solution
of a SDE have been established, we can analyse
the distribution of its solution at all time instants.
In describing the evolution of the distribution of a diffusion process,
the Fokker--Planck partial differential equation is a fundamental tool.
We therefore introduce the following assumption.
\begin{itemize}
 \item[(A2)]
The unique solution $X_t$ of ~(\ref{sde1}) admits a density $p_t$ that is absolutely
continuous with respect to the Lebesgue measure, i.e.,
\begin{eqnarray*}
\mbox{Prob}\{X_t\in A\} = \int_A p_t(x) dx, \ \ \mbox{for all Borel sets} \ \ A,
\end{eqnarray*}
and that satisfies the Fokker--Planck equation:
\begin{eqnarray} \label{FPESQRT}
\frac{\partial p_t}{\partial t} =  -\frac{\partial}{\partial x} (f_t p_t) + \half
   \frac{\partial^2}{\partial x^2} (a_t p_t),  \ \    a_t(\cdot) = \sigma_t^2(\cdot) \ .
\end{eqnarray}
\end{itemize}
Examples of assumptions on the coefficients $f$, $a$ and on their
partial derivatives
under which (A2) holds are given in the literature. See for example
Stroock and Varadhan (1979)
or Friedman (1975).

In order to  appropriately introduce the problem we mentioned at the beginning
of the section, we now present a definition of exponential family.
\begin{definition}
   Let $\{c_1,\cdots,c_m\}$ be scalar functions defined on ${\bf R}$,
   such that $\{1,c_1,\cdots,c_m\}$ are {\em linearly independent},
   have at most polynomial growth, are
   twice continuously differentiable and the convex set
\begin{displaymath}
   \Theta_0 := \left\{\theta=\{\theta^1,\ldots,\theta^m\}'\in {\bf R}^m\,:\,
   \psi(\theta) = \log\; \int \exp[ \theta' c(x) ]\, d x
   < \infty \right\}\ ,
\end{displaymath}
   has {\em non--empty interior}, where $c(x)=\{c_1(x),\cdots,c_m(x)\}'$ and
   `` $'$ '' denotes transposition.
   Then
\begin{displaymath}
   EM(c) = \{ p(\cdot,\theta)\,,\, \theta \in \Theta \},
   \hspace{1cm} p(x,\theta):= \exp[\theta' c(x) - \psi(\theta)]\ ,
\end{displaymath}
   where $\Theta \subseteq \Theta_0$ is open,
   is called an exponential family of probability densities.
\end{definition}

Our problem consists in finding a SDE whose solution $X_t$
has a density $p_t$ that follows a prescribed curve in a given exponential family.
More precisely, we require the curve $t \mapsto p_t$, in the space
of all densities, to coincide with a given curve
$t \mapsto p(\cdot,\theta_t)$ in a given $EM(c)$.\footnote{In order to contain space
and notation the underlying geometric setup
is not fully developed here. We just say that the problem originated from the
use of differential geometry and statistics for the nonlinear filtering
problem. The reader interested in geometric aspects and other details is referred to
Brigo (1996),  Brigo, Hanzon and Le Gland (1999), or to the tutorial in Brigo (1999).}

This is formalized in the following.
\begin{problem} \label{fin-dim:pro}
Let be given an exponential family $EM(c)$, an initial density $p_0$
contained in $EM(c)$, and
a diffusion coefficient $a_t(\cdot) := \sigma^2_t(\cdot)$,
$t\ge 0$.
Let ${\cal U}(p_0,\sigma)$ denote the set of all drifts
$f$ such that $p_0$, $f$ and $\sigma$ and its related SDE~(\ref{sde1}) satisfy assumptions
(A1) and (A2). Assume ${\cal U}(p_0,\sigma)$ to be non-empty.

Then, given the curve $t \mapsto p(\cdot,\theta_t)$ in $EM(c)$
(where $t \mapsto \theta_t$ is a $C^1$--curve in the parameter space
$\Theta$), find a drift in ${\cal U}(p_0,\sigma)$ whose related SDE has a solution
with density $p_t = p(\cdot,\theta_t)$.

\end{problem}
The solution of this problem is given by the following.
\begin{theorem} \label{sol-prob1}
{\bf (Solution of Problem~\ref{fin-dim:pro})}
Assumptions and notation of Problem~\ref{fin-dim:pro} in force.
Consider the stochastic differential equation
\begin{eqnarray} \label{sol:prob1}
d Y_t &=& u^\sigma_t(Y_t) dt + \sigma_t(Y_t) dW_t, \ \ Y_0 \sim p_0, \nonumber \\  \\
 u^\sigma_t(x) &:=& \half \frac{\partial a_t}{\partial x}(x) +
      \half a_t(x) \theta_t' \frac{\partial c}{\partial x}(x)
      \nonumber \\ \nonumber \\ \nonumber
&& - \left(\frac{d}{dt}\theta_t' \right)
         \int_{-\infty}^x \left(c(\xi) - \nabla_{\theta}\psi(\theta_t)\right)
         \ \exp[\theta_t' (c(\xi) - c(x))] d\xi,
\end{eqnarray}
where $\nabla_{\theta}\psi(\theta_t)=\{\partial \psi / \partial
\theta^1(\theta_t),\ldots,\partial \psi /\partial \theta^m(\theta_t) \}'$,
with the symbol ``$\sim$'' to be read as ``distributed as''.

If  $u^\sigma \in {\cal U}(p_0,\sigma)$,
then the SDE (\ref{sol:prob1}) solves Problem~\ref{fin-dim:pro}, in that
\begin{displaymath}
p_{Y_t}(x) = \exp\left[\theta_t' \ c(x) - \psi(\theta_t)\right],
\ \ t \ge 0.
\end{displaymath}
\end{theorem}
The proof of the theorem is rather straightforward.
It is sufficient to write the
Fokker--Planck equation for the SDE (\ref{sol:prob1}) and, after
lengthy computations, verify that indeed
\begin{displaymath}
 \frac{\partial}{\partial t} \exp[\theta_t' c(x) - \psi(\theta_t)]\
= - \frac{\partial}{\partial x}
\left(u^\sigma_t(x) \exp[\theta_t' c(x) - \psi(\theta_t)]\right)
+ \half  \frac{\partial^2}{\partial x^2}
\left( a_t(x) \exp[\theta_t' c(x) - \psi(\theta_t)] \right)  \
\end{displaymath}
by substituting the expression for $u$ given in the theorem.
A different proof can be found in Chapter 7 of Brigo (1996) or in Brigo (2000),
where in deriving the expression for $u$ it was tacitly assumed, as is done here, that
\begin{displaymath}
\lim_{x\rightarrow -\infty}u^\sigma_t(x)p_t(x) = 0 \ \ \mbox{for all } \ t \ge 0.
\end{displaymath}

In the next section, we shall consider an interesting application of this theorem to
the option pricing problem. Indeed, we shall use such result
more as a  ``guiding tool'' rather than applying it immediately
as it stands. In particular, assumptions (A1) and (A2) will be checked
directly and not via the sufficient conditions usually considered in the literature.

\section{Alternatives to the Black and Scholes model}
Let us consider the Black and Scholes (1973) stock price model,
\begin{eqnarray} \label{BeS}
 d S_t = \mu S_t dt + \sigb  S_t \ dW_t, \ \  S_0 = s_0 , \ \
t \in [0,T],
\end{eqnarray}
where $s_0$ is a positive deterministic initial condition, and $\mu$,
$\sigb$ and $T$ are positive real constants.

The probability density $p_{S_t}$ of $S_t$, at any time $t>0$,
is given by
\begin{eqnarray} \label{BeSlnd}
p_{S_t}(x) &=& \exp\left\{\zeta \ln\frac{x}{s_0} + \rho(t)
  \ln^2\frac{x}{s_0} - \psi(\zeta,\rho(t))  \right\}, \ \ x > 0, \\ \nonumber \\ \nonumber
\zeta &=& \frac{\mu}{\sigb^2} - \frac{3}{2},
\ \ \rho(t) = - \frac{1}{2 \sigb^2 t}, \\ \nonumber \\ \nonumber
\psi(\zeta,\rho(t)) &=& - \frac{(\zeta+1)^2}{4 \rho(t)}
+ \half \ln\left(\frac{-\pi}{\rho(t)}\right) + \ln(s_0).
\end{eqnarray}
With the notation for exponential families introduced in the previous section,
one writes
\begin{eqnarray*}
c_1(x) =  \ln\frac{x}{s_0}, \ \ c_2(x) =  \ln^2\frac{x}{s_0} , \ \
\theta_t = \{\zeta, \ \rho(t)\}' \ .
\end{eqnarray*}

One might wish to model the stock price process by considering a different volatility
function $\sigma$, instead of $\sigb S_t$ in (\ref{BeS})\footnote{We use
the term ``volatility'' to denote the whole diffusion coefficient $\sigma_t(\cdot)$ rather
than the standard deviation rate of the instantaneous return as usually done in practice.},
while preserving major properties of the original process
(\ref{BeS}). The purpose of this section is then the construction of
alternative stock price dynamics that differs from (\ref{BeS}), yet sharing
similar features from a probabilistic point of view.

Let us approach this problem by applying Theorem~\ref{sol-prob1}
to find a SDE with a given diffusion
coefficient $\sigma_t(\cdot)$ and whose marginal density is
equal to the marginal density of $S$ in all time instants
of the time interval ${\cal T} = [\epsilon,T]$, where $0<\epsilon<T$ and $\epsilon$
can be chosen arbitrarily close to $0$.
We remark that the density of $S$ is not concentrated in the whole
real line, but in $(0, +\infty)$.
However, the procedure resulting in Theorem~\ref{sol-prob1}
can be adapted in a straightforward way to the latter case
once $-\infty$ is replaced by $0$.

It easily follows from (\ref{sol:prob1}) that the equation sought for is
\begin{eqnarray} \label{sol:bes1}
d Y_t &=& u^\sigma_t(Y_t,s_0,0) dt + \sigma_t(Y_t) dW_t,
\ \ Y_\epsilon = S_\epsilon,
\ \  \epsilon \le t \le T, \nonumber \\  \\
\nonumber
 u^\sigma_t(x,y,\alpha) &:=& \half \frac{\partial a_t}{\partial x}(x) +
\half \frac{a_t(x)}{x} \left[ \zeta + 2 \rho(t-\alpha)
\ln\frac{x}{y}\right] +\frac{x}{2(t-\alpha)}
\left[\ln\frac{x}{y} - \frac{\zeta+1}{2 \rho(t-\alpha)}\right].
\end{eqnarray}
The definition of $Y$ is then extended to the whole interval
$[0, T]$ by setting
\[ d Y_t = \mu Y_t \ dt + \sigb Y_t dW_t, \ 0<t<\epsilon, \  Y_0 = s_0 \ . \]
In other terms, $Y$ is assumed to follow the same dynamics as the original
process $S$ in $[0,\epsilon)$, and to start from the same initial condition.

The reason for splitting the interval $[0,T]$ into two subintervals
is simply to avoid problems concerning the definition of the drift at time $t=0$.
The variable $\alpha$ is here introduced to allow for subsequent generalizations.

In general, we cannot prove that (\ref{sol:bes1}) satisfies assumptions
(A1) and (A2) for a prescribed $\sigma$, i.e. that $u^\sigma \in {\cal U}(p_0,\sigma)$
so that ${\cal U}(p_0,\sigma)$ is nonempty. Therefore, at this stage,
(\ref{sol:bes1}) is simply providing us with a {\em candidate} solution
of our problem for a generic $\sigma$.

In the following table we report some possible choices for $\sigma$
and the corresponding candidate $u^\sigma$.
In particular, we can notice that for $\sigma_t(x) = \sigb x$, $u^\sigma_t(x,s_0,0)$ equals,
as it must be, the original drift $\mu x$ of the Black and Scholes model (\ref{BeS}).

\begin{table}[ht]
\begin{center}
{\bf Table 1}\\
Examples of volatility functions and corresponding candidate drift
\mbox{} \\
\begin{tabular}{|c|c|}
\hline $\mbox{Volatility function} \ \ \sigma_t(x)$ & $ \mbox{Drift} \ \ u^\sigma_t(x,s_0,0)$
\\ \hline
$\sbar$ & $ \half \frac{\sbar^2}{x}
\left[ \zeta + 2 \rho(t) \ln\frac{x}{s_0}\right] +
\frac{x}{2t} \left[\ln\frac{x}{s_0} - \frac{\zeta+1}{2 \rho(t)}\right]$
\\
\hline $\sbar \sqrt{x}$  & $\frac{\sbar^2}{2}
(\frac{\mu}{\sigb^2}-\inverse{2}) -
\inverse{2t} \frac{\sbar^2}{\sigb^2} \ln\frac{x}{s_0} +
\frac{x}{2}(\mu - \frac{\sigb^2}{2}) +  \frac{x}{2t} \ln\frac{x}{s_0}$
\\
\hline $\sbar x$  & $ x \left[ \inverse{4}(\sbar^2 - \sigb^2) +
\frac{\mu}{2}(\frac{\sbar^2}{\sigb^2} + 1) \right]
+ \frac{x}{2t}\ln\frac{x}{s_0}(1-\frac{\sbar^2}{\sigb^2} )$  \\
\hline $\sigb x$ & $\mu x$ \\ \hline
\end{tabular} \\
\end{center}
\end{table}
Let us now assume that the
particular $\sigma$ we are working with is such that
$u^\sigma \in {\cal U}(p_0,\sigma)$.

We have seen so far that, by applying the results of the previous section,
it is rather straightforward to produce processes with the same marginal
distribution as that of (\ref{BeS}). However, a further fundamental
property of process (\ref{BeS}) is that its log-returns
between any two time instants are normally distributed,
independently of the prices at the considered instants, i.e.,
\begin{eqnarray} \label{BeSret}
 \ln \frac{S_{t+\delta}}{S_t}
\sim {\cal N}\left((\mu - \half \sigb^2)\delta ,\ \sigb^2 \delta\right),
\ \ \delta > 0, \ \ t\in [0,T-\delta].
\end{eqnarray}

Alternative models such as (\ref{sol:bes1}) do not share this
property in general. In fact, identity of the marginal laws alone
does not suffice to ensure (\ref{BeSret}), for which equality of
second order laws or of transition densities would be sufficient instead.
How can we obtain  alternative models whose properties concerning log-returns
are as close as possible to property~(\ref{BeSret})?

To tackle this issue, we have to find a compromise between our alternative
model (\ref{sol:bes1}) and model (\ref{BeS}). To this end, we consider a
weaker version of (\ref{BeSret}) in that we restrict the set of dates for
which the property holds true. Precisely, we modify the
definition of $Y$ so that, given the time instants
${\cal T}^\Delta:=\{0,\Delta, 2\Delta, \ldots, N\Delta\}$,
$\Delta = T/N$, $\Delta > \epsilon$,
property (\ref{BeSret}) is satisfied by $Y$ in ${\cal T}^\Delta$, i.e.
\begin{eqnarray} \label{yret}
 \ln \frac{Y_{i \Delta}}{Y_{j \Delta}}
\sim {\cal N}((\mu - \half \sigb^2)(i-j) \Delta ,\ \sigb^2(i-j) \Delta),
\ \ i > j, \ \ i=1,\ldots,N, \ \ j=0,\ldots,N-1.
\end{eqnarray}
Limiting such key property to a finite set of times is not so
dramatic. Indeed, only discrete time samples are observed in practice,
so that once the time instants are fixed, our  process $Y$ can not be
distinguished from Black and Scholes process'. The issue of discrete versus
continuous time will be further developed in Section 5.

The new definition of $Y$ is still based on Theorem~\ref{sol-prob1}.
However, we use this theorem ``locally'' in each time interval $[(i-1)\Delta, \ i\Delta)$.
This means that in such interval we define iteratively the drift $u^\sigma$ as in the
theorem but
\begin{itemize}
\item we translate back the time--dependence of a time amount $(i-1)\Delta$
     (thus locally restoring the dynamics of the original result) and
\item we replace the distribution $p_0$ for the initial condition with
      the distribution of the final value of $Y$ relative to the previous interval.
\end{itemize}
We obtain:
\begin{eqnarray}
\label{sol:bes2}
d Y_t &=& u^\sigma_t(Y_t,Y_{\alpha(t)},\alpha(t)) dt +  \sigma_t(Y_t) dW_t, \ \
t \in [i\Delta + \epsilon, (i+1)\Delta),  \\ \nonumber \\ \nonumber
dY_t &=& \mu Y_t dt + \sigb Y_t dW_t,  \ \ \mbox{for} \ \  t \in [i\Delta,i\Delta+ \epsilon),\ \
\alpha(t) = i \Delta \ \ \mbox{for} \ \  t \in [i\Delta, \ (i+1)\Delta) \ ,
\end{eqnarray}
where $u^\sigma_t(x,y,\alpha)$ was defined in (\ref{sol:bes1}).

It is clear by construction that the transition densities
of $S$ and $Y$ satisfy
$p_{Y_{(i+1) \Delta}|Y_{i \Delta}}(x;y) = p_{S_{(i+1) \Delta}|S_{i\Delta}}(x;y)$.
Then, starting from the equality of the marginal laws of $S$ and $Y$ in the first
interval that holds by construction, we inductively obtain the equality of the
marginal laws also in each other interval.
As a consequence, the second order densities are also equal
among consecutive instants $(i-1)\Delta, \ i\Delta$, i.e.,
\begin{displaymath}
p_{[Y_{(i+1) \Delta},Y_{i\Delta}]}(x,y) =
p_{[S_{(i+1) \Delta},S_{i\Delta}]}(x,y).
\end{displaymath}
It follows that
\begin{eqnarray} \label{yret1}
 \ln \frac{Y_{(i+1) \Delta}}{Y_{i \Delta}}
\sim {\cal N}((\mu - \half \sigb^2) \Delta ,\ \sigb^2 \Delta),
\ \ i=0,\ldots,N-1.
\end{eqnarray}
At this point we remark that the process $Y$ in (\ref{sol:bes2}) is not a Markov process
in $[0, T]$. However, it is Markov in all time instants of ${\cal T}^\Delta$.
Formally,
\begin{displaymath}
p_{Y_{m \Delta}|Y_{(m-1)\Delta},Y_{(m-2)\Delta},\ldots,Y_0} =
p_{Y_{m \Delta}|Y_{(m-1)\Delta}}.
\end{displaymath}
This property follows from the fact that in $[(m-1)\Delta, m \Delta)$
the dynamics of the SDE defining $Y$  does not depend on
$Y_{(m-2)\Delta},\ldots,Y_0$, and that when such equation
is considered for $t\in [(m-1)\Delta, m\Delta)$, in its drift $u^\sigma$
the local initial condition for the entry $Y$ is set to $Y_{(m-1)\Delta}$.

From now on, we refer to markovianity in  ${\cal T}^\Delta$ as to
$\Delta${\em --Markovianity}.

We finally notice that, through the $\Delta$--Markovianity,
property (\ref{yret1}) extends to any pair of instants in
${\cal T}^\Delta$, so as to yield (\ref{yret}).
Moreover, the inductive application of the $\Delta$--Markovianity
and the identity of transition densities in the grid leads to the
identity of the finite dimentional distributions of $S$ and $Y$ in the grid.

\section{Option pricing in continuous-time} \label{ophct}
Let us now consider the process $\{B_t:t\ge 0\}$ whose value evolves
according to
\begin{equation}
dB_t=B_t r dt,
\end{equation}
with $B_0=1$ and where $r$ is a positive real number, so that
$B_t = \exp(r t)$. The process $B$ is assumed to describe the evolution of
a money market account in a given financial market. The process $Y_t$ in
(\ref{sol:bes2}) is instead assumed to model the evolution of some traded
financial (risky) asset, typically a stock.

The financial market thus defined might admit arbitrage opportunities. As
is well known, a sufficient condition which ensures
arbitrage-free dynamics is the existence of an equivalent martingale measure
with respect to the initially chosen numeraire. In this paper, we use the
process $B$ as a numeraire, so that an equivalent martingale measure is a
probability measure that is equivalent to the initial one, $P$, and under
which the process $\{Y_t/B_t:t\ge 0\}$ is a martingale. A necessary condition for
the existence of an equivalent martingale measure is the semimartingale
property for the process $Y$. The process $Y$ is indeed a semimartingale
under $P$ for sufficiently well behaved volatility functions $\sigma_t(\cdot)$.

We denote by $\cal S$ the set of all volatility functions $\sigma_t(\cdot)$
such that $u^\sigma \in {\cal U}(p_0,\sigma)$ and for which there exists a unique
equivalent martingale measure.

The set $\cal S$ is obviously non-empty, since it contains at least the
Black and Scholes volatility function $\sigma_t(x)=\sigb x$. Moreover, as we will
prove in the sequel, all volatilities functions of type $\sbar I$, $\sbar >0$, belong
to $\cal S$, with $I$ denoting the identity map. An interesting example of
volatility functions which do not belong to ${\cal S}$ is instead provided in the appendix.

We now assume that we have chosen $\sigma\in {\cal S}$ and the
corresponding equivalent martingale measure $Q^\sigma$.
Since $\{Y_t/B_t:t\ge 0\}$ is a martingale under such a measure, it easily
follows that under $Q^\sigma$ the process $Y$ satisfies the SDE
\begin{eqnarray*}
d Y_t &=& r Y_t \ dt + \sigb Y_t \ d \widetilde{W}_t,  \ \  t \in [i\Delta, i\Delta+\epsilon), \\
d Y_t &=& r Y_t \ dt + \sigma_t(Y_t) \ d \widetilde{W}_t,  \ \ t \in [i\Delta+\epsilon, (i+1)\Delta),
\end{eqnarray*}
where  $\widetilde{W}$ is a standard Brownian motion under
$Q^\sigma$.

Furthermore, under the assumption that
i) there are no-transaction costs, ii) the borrowing and lending rates are both equal to $r$, iii)
short selling is allowed with no restriction or penalty, and iv)
the stock is infinitely divisible and pays no dividends,
the unique no-arbitrage price for a given contingent claim
$H\in L^2(Q^\sigma)$ is (see Harrison and Pliska (1981, 1983))
\begin{equation}
\label{optpr}
V_t=\frac{B_t}{B_T}E^{Q^\sigma}\left\{\left.H\right| {\cal F}_t \right\},
\end{equation}
where $\{{\cal F}_t:t\ge 0\}$ denotes the filtration associated to the
process $Y$.

In the special case of a European call option, the following are
interesting problems to solve.
\begin{problems} \label{infsup}
Let us assume that the given claim is a European call option, written on
the stock, with maturity $T$ and strike $K$. Find:
\begin{equation}
\label{optsiginf}
\inf_{\epsilon>0,\sigma\in {\cal S}} B_T^{-1}
E^{Q^\sigma}\left\{\left.(Y_T-K)^+\right| {\cal F}_0 \right\},
\end{equation}
\begin{equation}
\label{optsigsup}
\sup_{\epsilon>0,\sigma\in {\cal S}} B_T^{-1}
E^{Q^\sigma}\left\{\left.(Y_T-K)^+\right| {\cal F}_0 \right\}.
\end{equation}
\end{problems}
Solving these problems is equivalent to finding the lowest and highest
theoretical price of the option for which the underlying stock price has lognormal
marginal distribution and normal log-returns on the grid ${\cal T}^\Delta$, with
standard deviations proportional to $\sigb$.

If we denote by $V_{*}$ the value of the infimum in
(\ref{optsiginf}),
and by $V^{*}$ the value of the supremum in (\ref{optsigsup}),
the following inequalities obviously hold:
\begin{equation}
\label{ineqbs}
 (s_0 - Ke^{-rT})^+ \le V_{*}\leq V_{BS}(\sigb)
 \le V^{*} \le s_0,
\end{equation}
where $V_{BS}(\sigb)$ denotes the option Black and Scholes price at time 0
as determined by $\sigb$, the volatility parameter in (\ref{BeS}).
Indeed, since $\sigb I\in {\cal S}$, the central inequalities hold by
definition of $V_{*}$ and $V^{*}$, whereas the first and
the last ones feature respectively the well known no-arbitrage lower and upper
bounds for option prices.

In the next subsection we will show that the first and last inequalities in
(\ref{ineqbs}) are actually equalities. To prove this statement, it will be
sufficient to restrict our analysis to the class of volatilities
$\{\sigma_t(x) = \sbar x$, $\sbar > 0\}$. This result is at first sight surprising.
Indeed, one would naively expect that the difference
between prices implied by models which are equivalent in the $\Delta$--grid
is bounded by a quantity that is somehow related to $\Delta$, typically
${\cal O}(\Delta^\lambda)$ for some positive real $\lambda$. In fact, by
halving the size of $\Delta$, we double the discrete--time instants where
the models in our family are equivalent. Accordingly, we would expect the prices
implied by these now ``closer'' models to range in a narrower interval. However, as
we shall soon see, this is not the case.

\subsection{A fundamental case} \label{fundamental}
We begin by stating the following.
\begin{lemma}
In the fundamental case where $\sigma_t(x) = \sbar x$, $\sbar>0$, the
process $Y^\sbar$ given by
\begin{eqnarray} \label{sol:bes3}
d Y_t^\sbar &=& u^{\sbar I}_t(Y_t,Y_{\alpha(t)},\alpha(t)) dt + \sbar \ Y_t \ dW_t,
\ \ t \in [i\Delta + \epsilon, (i+1)\Delta), \nonumber \\ \\ \nonumber
u^{\sbar I}_t(y,y_\alpha,\alpha) &=&
y \left[ \inverse{4}(\sbar^2 - \sigb^2) +
\frac{\mu}{2}(\frac{\sbar^2}{\sigb^2} + 1) \right]
+ \frac{y}{2(t-\alpha)} (1-\frac{\sbar^2}{\sigb^2} ) \ln\frac{y}{y_\alpha},
\\ \nonumber \\ \nonumber
 d Y_t^\sbar &=& \mu Y_t^\sbar dt + \sigb Y_t^\sbar dW_t \ \ \mbox{for} \ \  t \in [i\Delta,i\Delta+ \epsilon),\ \
\alpha(t) = i \Delta \ \ \mbox{for} \ \  t \in [i\Delta, \ (i+1)\Delta) \
\end{eqnarray}
solves Problem~\ref{fin-dim:pro} when $p_t$ is given by (\ref{BeSlnd}).
Moreover, the volatility function $\sigma_t(x) = \sbar x$ belongs to $\cal S$,
for any $\sbar>0$.
\end{lemma}
{\em Proof}.
Since $Y^\sbar$ has the same marginal distribution
as $S$ under $P$, it follows that $Y_t^\sbar>0$. Then, the process $Z_t = \ln Y_t^\sbar$ is
well defined and, by It\^o's formula, the SDE for $Z_t$ is piecewise linear in a narrow
sense (in that its diffusion coefficient is purely deterministic), and hence
admits a unique strong solution which, for each $t\in [j\Delta,(j+1)\Delta)$,
is explicitly given by
\begin{eqnarray} \label{intediff}
Z_t &=& Z_{j\Delta} + (\mu -\half\sigb^2)(t-j\Delta)  \\ \nonumber
& &
+\left\{
\begin{array}{ll}
\sigb(W_t-W_{j\Delta})  & t\in [j\Delta,j\Delta+\epsilon), \\ \nonumber
\left(\frac{t-j\Delta}{\epsilon}\right)^{\beta/2}
\left[\sigb(W_{j\Delta+\epsilon}-W_{j\Delta})+ \sbar\int_{j\Delta+\epsilon}^t
\left(\frac{u-j\Delta}{\epsilon}\right)^{-\beta/2} dW_u\right]
& t\in [j\Delta+\epsilon,(j+1)\Delta),
\end{array}
\right.
\end{eqnarray}
where $\beta = 1-\frac{\sbar^2}{\sigb^2}$.

As a consequence, the assumptions of Problem~\ref{fin-dim:pro} are satisfied
so that $u^{\sbar I}$ solves Problem~\ref{fin-dim:pro} when $p_t$ is given by
(\ref{BeSlnd}). Moreover, the Girsanov change of measure from $P$ to $Q^{\sbar I}$
is well defined since one can show that the Novikov condition is satisfied through
application of the ``tower property'' of conditional expectations.
Hence, the measure $Q^{\sbar I}$ exists unique and $\sbar I$ belongs to
$\cal S$ for each $\sbar > 0$. For the uniqueness of the measure $Q^{\sbar I}$,
we refer for instance to Duffie (1996).

\begin{theorem}
In the fundamental case where $\sigma_t(x) = \sbar x$, $\sbar>0$,
the unique no-arbitrage option price at time $t$ is the Black and Scholes price
\begin{equation}
\label{bslem}
U^\epsilon(t,\sbar)=Y_t^\sbar \Phi(d_1) - K e^{-r(T-t)}\Phi(d_2),
\end{equation}
where
\begin{eqnarray*}
&& d_1=\frac{\ln(Y_t^\sbar/K)+(r+\nub^\epsilon (t)^2/2)(T-t)}{\nub^\epsilon (t) \sqrt{T-t}},\\ \\
&& d_2=d_1(t) - \nub^\epsilon (t) \sqrt{T-t},
\end{eqnarray*}
\begin{equation}
\label{nubar}
\nub^\epsilon (t)=
\left\{
\begin{array}{ll}
\sqrt{\frac{
\frac{\epsilon}{\Delta}(\sigb^2-\sbar^2)(T-\alpha(t))+\sbar^2(T-\alpha(t))
+\sigb^2(\alpha(t)-t)}{T-t}} & t \in [\alpha(t),\alpha(t)+\epsilon)
\\ \sqrt{\frac{
\frac{\epsilon}{\Delta}(\sigb^2-\sbar^2)(T-\alpha(t)-\Delta)+
\sbar^2(T-t)}{T-t}} & t \in [\alpha(t)+\epsilon,\alpha(t)+\Delta)
\end{array}
\right.
\end{equation}
\end{theorem}
{\em Proof}.
From the previous lemma, we infer the existence of a unique
no-arbitrage option price that can be calculated through (\ref{optpr}).
Then (\ref{bslem}) is obtained by noticing that under the equivalent martingale measure
\[ \ln \frac{Y_T^\sbar}{Y_t^\sbar}
\sim {\cal N}\left((r - \half \nub^\epsilon (t)^2)(T-t) ,\ \nub^\epsilon
(t)^2 (T-t)\right), \ t\in [0,T],\]
with $\nub^\epsilon$ given by (\ref{nubar}). This implies that the option price
at time $t$ corresponding to $Y^\sbar$ is the Black and Scholes price with
volatility coefficient $\nub^\epsilon(t)$, i.e., that (\ref{bslem}) holds.

\mbox{}\newline
The key point of our result is that, for any given volatility
coefficient $\sbar$, we are free to adjust the drift of the SDE defining the dynamics of
the stock--price process under the objective measure,
in such a way that the resulting $Y^\sbar$ has the same distributional properties of
the Black and Scholes process on discrete-time dates. As opposed to this,
the risk--neutral valuation for pricing options {\em imposes the drift} $r Y^\sbar$ to
the SDE followed by $Y^\sbar$ under the equivalent martingale measure.
This causes the option price implied by the alternative model to coincide, at first order
in $\epsilon$, with the Black and Scholes price with volatility parameter $\sbar$. In fact,
imposing the drift $r Y^\sbar$ to $Y^\sbar$ leads to the same risk neutral process as that
of Black and Scholes' (obtained by imposing the drift $rS$ to $S$), with the only difference
that $\sigb$ is replaced by $\sbar$.

\begin{remark} {\bf (Historical versus Implied volatility).}
A first interesting property that can be deduced from this theorem
applies in case one believes that option prices trade independently of the
underlying stock price.  We have in fact been able to
construct a stock price process, the process (\ref{sol:bes3}), whose marginal
distribution and transition density depend on the volatility coefficient $\sigb$,
whereas the corresponding option price, in the limit for $\epsilon \rightarrow 0$,
only depends on the volatility coefficient $\nub$. As a consequence, we can  provide a
consistent theoretical framework which justifies the differences between
historical and implied volatility that are commonly observed in real markets.
\end{remark}

Straightforward application of the previous theorem leads to the main result
of this section which is summarized in the following.
\begin{corollary} \label{inf}
The solutions of problems
(\ref{optsiginf}) and (\ref{optsigsup}) are
\begin{eqnarray}
\label{solinf}
&V_{*}=(s_0-Ke^{-rT})^+\nonumber \\ \\
&V^{*}=s_0.\nonumber
\end{eqnarray}
Moreover, for any other candidate price $\bar{V} \in (V_{*},V^{*})$
there exist a volatility $\sbar$ and an $\epsilon > 0$ such that
\begin{eqnarray*}
U^\epsilon(0,\sbar) = \bar{V} .
\end{eqnarray*}
\end{corollary}
{\em Proof}.
To prove (\ref{solinf}), we simply have to take the limit of expression
(\ref{bslem}) (with $t=0$) for $\epsilon$ going to zero and $\sbar$ either going to zero
or going to infinity, since
\begin{eqnarray*}
&&\lim_{v \rightarrow 0} V_{BS}(v) = (s_0 - Ke^{-rT})^+, \\
&&\lim_{v \rightarrow +\infty} V_{BS}(v) = s_0,
\end{eqnarray*}
and
\[ \lim_{\epsilon\rightarrow 0} \nub^\epsilon (0)=\sbar.\]
Finally, we remember that, ceteris paribus, $V_{BS}(v)$ is a strictly increasing
function of $v$, so that $V_{BS}(v)=\hat{V}$ has a unique solution for
$V \in (V_{*},V^{*})$, hence $U^\epsilon(0,\sbar) = \bar{V}$ has a solution in
$(0,+\infty)\times (V_{*},V^{*})$.

\begin{remark} {\bf (Taking $\epsilon \rightarrow 0$ ).}
It is possible to consider the limit for $\epsilon \rightarrow 0$ in the above
expressions so as to present our result in a simpler and more elegant way.
However, the treatment with $\epsilon$ does not involve limit considerations
and permits to contain analytical effort, so that we decided to keep $\epsilon > 0$.
This can be also useful in numerical implementations.
We just observe that, for $\epsilon \rightarrow 0$, since $\beta<1$,
(\ref{intediff}) becomes
\begin{eqnarray*}
Z_t &=& Z_{j\Delta} + (\mu -\half\sigb^2)(t-j\Delta)  \\ \nonumber
& & +
 (t-j\Delta)^{\beta/2}  \sbar\int_{j\Delta}^t
 (u-j\Delta)^{-\beta/2} dW_u
\ \ \  t\in [j\Delta ,(j+1)\Delta).
\end{eqnarray*}
This process is well defined since the integral in the right-hand side exists finite a.s.
even though its integrand diverges when $u \rightarrow j \Delta^+$.

The above equation can be better compared to the Black and Scholes
process when written in differential form:
\begin{eqnarray*}
d Z_t = (\mu -\half\sigb^2)\ dt +
\frac{\beta}{2}(t-j\Delta)^{\beta/2-1}  \sbar\int_{j\Delta}^t (u-j\Delta)^{-\beta/2} dW_u\ dt
  + \sbar \ dW_t  \ \  \  t\in [j\Delta ,(j+1)\Delta).
\end{eqnarray*}
By observing this last equation we can isolate three terms in the right-hand side.
The first term is the same drift as in the log--returns of the Black and Scholes
process. The third term is the same as in the log--returns of the Black and Scholes
process, but the volatility parameter $\sigb$ is replaced by our $\sbar$.
Finally, the central term is the term which is needed to have returns equal to
the returns in the Black and Scholes process even after changing the volatility
from $\sigb$ to $\sbar$. Note that this term goes to zero for $\sigb = \sbar$.
It is this term that makes our process non-Markov outside the trading time grid.
\end{remark}

The interpretation of the previous theorem and corollary is as
follows. If we are given a discretely observed stock price, the particular
way we use to ``complete'' the model with any of our continuous-time processes $Y$ has a
heavy impact on the associated option price. The influence is in fact so
relevant that such a price can be arbitrarily close to either no-arbitrage bounds
for option prices.

From an intuitive point of view, the reason why option prices can be so
different is because the time step in the grid ${\cal T}^\Delta$
is never infinitesimal. In other words, continuous-time option prices
would simply reveal the differences existing at an infinitesimal level
among all the stock price processes $Y^\sbar$.

\section{Option pricing and hedging in the real world}
Let us now consider a trader who needs to price an option on a given stock.
His usual practice is to resort to continuous-time mathematics to fully
exploit the richness of its theoretical results, and to model stock returns with
a normal distribution.

Let us denote by $\delta t$ the length of the smallest time interval
when an actual transaction can occur. Such $\delta t$ is the best realistic
approximation of the infinitesimal time distance ``$dt$''.

The results of the previous section imply that
the geometric Brownian motion (\ref{BeS}) is
just one of the infinitely many processes $Y$ that possess the
properties required by the practitioner along intervals of equal length $\delta t$.
However, the basic equivalence in the description of
the stock price dynamics can not be extended to the corresponding option
prices. Indeed, any real number in the interval $(V_{*},V^{*})$ can be
viewed as the unique no-arbitrage option price for some process $Y$.

Which process $Y$ should then be chosen by the practitioner?

A possible answer to this question can be provided,
for example, through the estimation of the option replication
error in discrete time.
To this end, for any $\sbar >0$, let us denote by $(\xi^\sbar,\eta^\sbar)$
the self-financing strategy that replicates in continuos time the option payoff for
the process $Y^\sbar$, where
$\xi_t^\sbar$ and $\eta_t^\sbar$ are respectively interpreted as the number of stock
shares and money units held at time $t$.


Then, fixing a set of dates $\tau_0=0<\tau_1<\cdots<\tau_n=T$ such that
$\tau_i=i \delta t$,\footnote{
Although fixed a priori, the set ${\cal T}^\Delta$
introduced earlier is arbitrarily chosen, so that we can set
$\Delta=\delta t/k$, with $k$ any positive integer implying that
$\{\tau_1,...,\tau_n\}  \subset {\cal T}^\Delta$. }
and, denoting the observed stock price at time $\tau_j$ by $\bar{S}_{\tau_j}$,
$j=0,\ldots,n$, the replication error when hedging according to the
strategy $(\xi^\sbar,\eta^\sbar)$, starting from the endowment $U^\epsilon(0,\sbar)$, is
\begin{eqnarray*}
\varepsilon(\sbar):=(\bar{S}_{T}-K)^+ -U^\epsilon(0,\sbar)-\sum_{j=0}^{n-1}
\xi^\sbar_{\tau_j}(\bar{S}_{\tau_{j+1}}-\bar{S}_{\tau_j})
-\sum_{j=0}^{n-1}\eta^\sbar_{\tau_j}(e^{r \tau_{j+1}}-e^{r \tau_j}).
\end{eqnarray*}
%
At this stage, we can solve, for example, an
optimization problem where $\varepsilon(\sbar)$ is minimized,
according to some criterion, over all $\sbar>0$.
Such procedure, however, is justified only to
measure the performance of our continuous-time prices and strategies on the fixed set
of discrete-time instants.
More generally, the issue of deriving a fair $\delta t$-time option price and hedging
strategy should be tackled by resorting to the existing literature on incomplete markets.

Many are the criterions one can choose from for pricing and hedging in
incomplete markets. We mention for instance those of
F\"{o}llmer and Sondermann (1986), F\"{o}llmer and Schweizer (1991),
Schweizer (1988, 1991, 1993, 1994, 1995, 1996), Sch\"{a}l (1994), Bouleau and
Lamberton (1989), Barron and Jensen (1990), El Karoui, Jeanblanc-Picqu\'{e}
and Viswanathan (1991), Davis (1994), Frasson and Runggaldier (1997),
El Karoui and Quenez (1995), Frittelli (1996), Mercurio (1996), Mercurio
and Vorst (1997), Bellini and Frittelli (1997), Frey (1998) and F\"ollmer and Leukert
(1998).

However, the purpose of this section is not to favor any particular approach.
We want, instead, to stress the following innovative feature in the theoretical
problem of option pricing. Instead of fixing a stock price process and then deriving
a fair option price and an ``optimal'' hedging strategy, we can in fact consider
a family of processes, that are somehow equivalent in the description of the
stock price evolution, among which we can select a convenient one by means of our
favorite incomplete markets criterion.


A comparison between the performances of the strategy $(\xi^\sbar,\eta^\sbar)$ and the hedging
strategy associated to any incomplete-market criterion is beyond the scope of the paper and is
left to future research.

\section{Conclusions}
In the present paper we consider an option-pricing application of a result which
is based on the construction of nonlinear SDE's with densities evolving
in a given finite-dimensional exponential family.
Precisely, we derive a family of stock price models that behave almost
equivalently to that of Black and Scholes.
All such models share the same distributions for the stock price process and
its log-returns along any previously fixed `trading time-grid'.
Therefore, all these models can be viewed as equivalent in the description of the stock
price evolution.

However, the continuous-time dynamics chosen to `complete' the model
in between the instants of the trading time-grid,
reflects heavily on the option price. The option price in fact can assume any value
between the option intrinsic value and the underlying stock price.

As a conclusion, our result points out that no dynamics is the {\em right}
one a priori, and that an incomplete-market criterion is needed to choose
among all the different models.
Practitioners with different criteria can still pick up a model
from our family, so as to match their expectations or to minimize their exposures.

Even though our results are based on the assumption of a lognormal distribution
for the stock price and a normal distribution for its log-returns, the generalization
to many other distributions is possible. Notice, indeed, that
the results of Section 2 can be applied to any curve of densities in a given
exponential family. However, the generalization is not straightforward and heavily relies
on the particular distributions which are considered.

\section*{Appendix}
In this appendix we consider the case $\sigma_t(\cdot) \equiv \sbar \neq 0$
as an interesting example of a class of volatilities which do not belong to ${\cal S}$.
Suppose, for the sake of simplicity,
that we require only the marginal law of $Y$ to coincide with
the marginal law of $S$, ignoring the returns distributions.
Now assume by contradiction that  $\sbar \in {\cal S}$.
Then by applying~(\ref{sol:bes1}) we obtain a Markov process~$Y$
\begin{eqnarray} \label{gaussy}
dY_t =  \half \frac{\sbar^2}{Y_t}
\left[ \zeta + 2 \rho(t) \log\frac{Y_t}{s_0}\right] \ dt +
\frac{Y_t}{2t} \left[\log\frac{Y_t}{s_0} - \frac{\zeta+1}{2 \rho(t)}\right]
\ dt + \sbar \ dW_t \ , \ \ \epsilon \le t \le T \ ,
\end{eqnarray}
with $Y_t=S_t$ for each $t \in [0,\epsilon]$.

Let us focus on $t \ge \epsilon$.
%
%
%
Since we now know that $Y_t$ and $S_t$ have the same
distribution under the objective probability measure, $Y_t$ is lognormally
distributed, and in particular $Y_t > 0$ for all $t$.
Since $\sbar \in {\cal S}$, there exists an equivalent martingale measure,
so that, under such a measure,
\begin{eqnarray*}
dY_t = r Y_t \ dt + \sbar \ d\widetilde{W}_t \ ,
\end{eqnarray*}
where $\widetilde{W}$ is a standard Brownian motion under
the martingale measure. Now notice that this last SDE is a linear
equation, so that its solution has a density whose
support is the whole real line. In other terms, such solution
can be negative with positive probability at any fixed
time instant, as opposed to what we have seen
for $Y$ under the objective measure. However, there cannot be an equivalent measure
that transforms a process whose support is the positive halfline
into another one whose support is the whole real line. Therefore, we
have contradicted our assumption that $\sbar \in {\cal S}$.

Our example is similar in spirit to that of Delbaen and Schachermayer (1995)
who consider a Bessel process (taking positive values at all times)
which cannot be transformed into a Brownian motion. Some conditions
for the existence of an equivalent martingale measure are given
for example in Rydberg (1997).

\end{document}